\documentclass{ckm}                 

\newcommand{\MeV}{\rm{MeV}}
\newcommand{\rhobar}{\bar {\rho}}
\newcommand{\etabar}{\bar{\eta}}
\newcommand{\dmd}{\Delta m_d}
\newcommand{\dms}{\Delta m_s}
\newcommand{\BK}{B_K}
\newcommand{\epsilonk}{\left|\varepsilon_K \right|}
\newcommand{\fbdsqbd}{F_{B_d} \sqrt{\hat B_{B_d}}}

\confname{Workshop on the CKM Unitarity Triangle, IPPP Durham, April
  2003}

\title{Unitarity Triangle Analysis in the Standard Model and Sensitivity to New 
Physics\thanks{Talk presented by V.~Lubicz}}

\author{M Ciuchini\addressmark{a}, E Franco\addressmark{b}, 
F Parodi\addressmark{c}, V Lubicz\addressmark{d,a}, 
L Silvestrini\addressmark{b} and A Stocchi\addressmark{e}}

\address[a]{INFN, Sezione di Roma III, Via della Vasca Navale 84, I-00146 Rome, 
Italy}
\address[b]{INFN, Sezione di Roma, P.le Aldo Moro 2, I-00185 Rome, Italy}
\address[c]{Dipartimento di Fisica, Universit\`a di Genova and INFN, Via
Dodecaneso 33, I-16146 Genova, Italy}
\address[d]{Dipartimento di Fisica, Universit\`a di Roma Tre, Via della Vasca 
Navale 84, I-00146 Rome, Italy}
\address[e]{LAL, IN2P3-CNRS et Universit\'e de Paris-Sud, BP 34, F-91898 Orsay 
Cedex, France}

\begin{document}

\begin{abstract}{
By using the most recent determinations of the several theoretical and
experimental input parameters, we update the Unitarity Triangle analysis in the
Standard Model and discuss the sensitivity to New Physics effects. We
investigate the interest of measuring with a better precision the various
physical quantities entering the Unitarity Triangle analysis and study in a
model independent way whether, despite the undoubted success of the CKM
mechanism in the Standard Model, the Unitarity Triangle analysis still allows
the presence of New Physics.}
\end{abstract}

\maketitle


\section{Introduction}
The analysis of the Unitarity Triangle (UT) and CP violation represents one of 
the most stringent tests of the Standard Model (SM) and, for this reason, also 
an interesting window on New Physics (NP). The input of this analysis is 
a large number of both experimental and theoretical parameters, the most 
relevant of which are listed in Table \ref{tab:inputs}. 
\begin{table*}[htb!]
\begin{center}
\begin{tabular}{|c|c|c|c|c|}
\hline Parameter & Value & Gaussian &  Theory \\
                &        & $\sigma$ & uncertainty  \\ \hline
$\lambda$ & 0.2240({\it 0.2210}) & 0.0036 ~({\it 0.0020}) & - \\ \hline
$\left | V_{cb} \right | (\times 10^{-3})$ (excl.) &  42.1  & 2.1 & - \\
$\left | V_{cb} \right | (\times 10^{-3})$ (incl.) &  41.4~({\it 40.4}) & 
        0.7 &  0.6({\it 0.8}) \\ \hline 
$\left | V_{ub} \right | (\times 10^{-4})$ (excl.) &  33.0({\it 32.5})  & 
        2.4({\it 2.9}) &  4.6({\it 5.5})  \\
$\left | V_{ub} \right | (\times 10^{-4})$ (incl.) &  40.9  &  
        4.6 &  3.6  \\ \hline
$\Delta M_d~(\mbox{ps}^{-1})$ & 0.503~({\it 0.494}) & 0.006~({\it 0.007}) & - \\
$\Delta M_s~(\mbox{ps}^{-1})$ & $>$ 14.4~({\it 14.9}) at 95\% C.L. & 
        \multicolumn{2}{|c|} {sensitivity 19.2~({\it 19.3})}  \\
$m_t$ (GeV) &  167  &  5  &  -   \\
$m_c$ (GeV) &  1.3  &  -  &  0.1 \\
$F_{B_d} \sqrt{\hat B_{B_d}}$(MeV)  &  223~({\it 230}) &   33~({\it 30}) & 
        12~({\it 15})  \\
$\xi=\frac{ F_{B_s}\sqrt{\hat B_{B_s}}}{ F_{B_d}\sqrt{\hat B_{B_d}}}$ & 
        1.24({\it 1.18})  &  0.04~({\it 0.03})  & 0.06~({\it 0.04}) \\ \hline
$\hat B_K$  &  0.86 &  0.06  &  0.14  \\ \hline
sin 2$\beta$  & 0.734~({\it 0.762}) & 0.054~({\it 0.064})  & -  \\ \hline
\hline
\end{tabular} 
\end{center}
\caption {\it {Values of the relevant quantities used in the fit of the CKM 
parameters. In the third and fourth columns the Gaussian and the flat parts of 
the uncertainty are given, respectively. All central values and errors are 
those adopted in ref.~\cite{yb}. The values within parentheses are the ones 
available at the time of the first CKM Workshop.}}
\label{tab:inputs} 
\end{table*}
A careful choice (and a continuous update) of the values of these parameters 
represents a crucial ingredient in this study, and this was indeed one of the 
main tasks of the first CKM Workshop. The conclusions of that Workshop have been
reported in ref.~\cite{yb}. In this talk we update the analysis of the UT in the
SM by assuming the central values and errors of the input parameters adopted 
in~\cite{yb} and collected in Table \ref{tab:inputs}.

The second part of this talk is dedicated to NP. We address two different 
(though related) questions. The first one concerns the interest of measuring the
various physical quantities entering the UT analysis with a better precision. We
investigate, in particular, to which extent future and improved determinations 
of the experimental constraints, sin2$\beta$, $\dms$ and $\gamma$, could allow 
us to invalidate the SM, thus signalling the presence of NP effects. The second 
question concerns the possibility of having significant NP contributions in the 
present analysis of the UT. Given the actual theoretical and experimental 
constraints, we investigate to which extent the UT analysis can still be 
affected by NP contributions. We show that, despite the undoubted success of 
the SM in describing the flavour sector, large NP contributions in this analysis
are in fact still possible (although unnecessary).
\begin{figure}[t]
\hbox to\hsize{\hss
\includegraphics[width=\hsize]{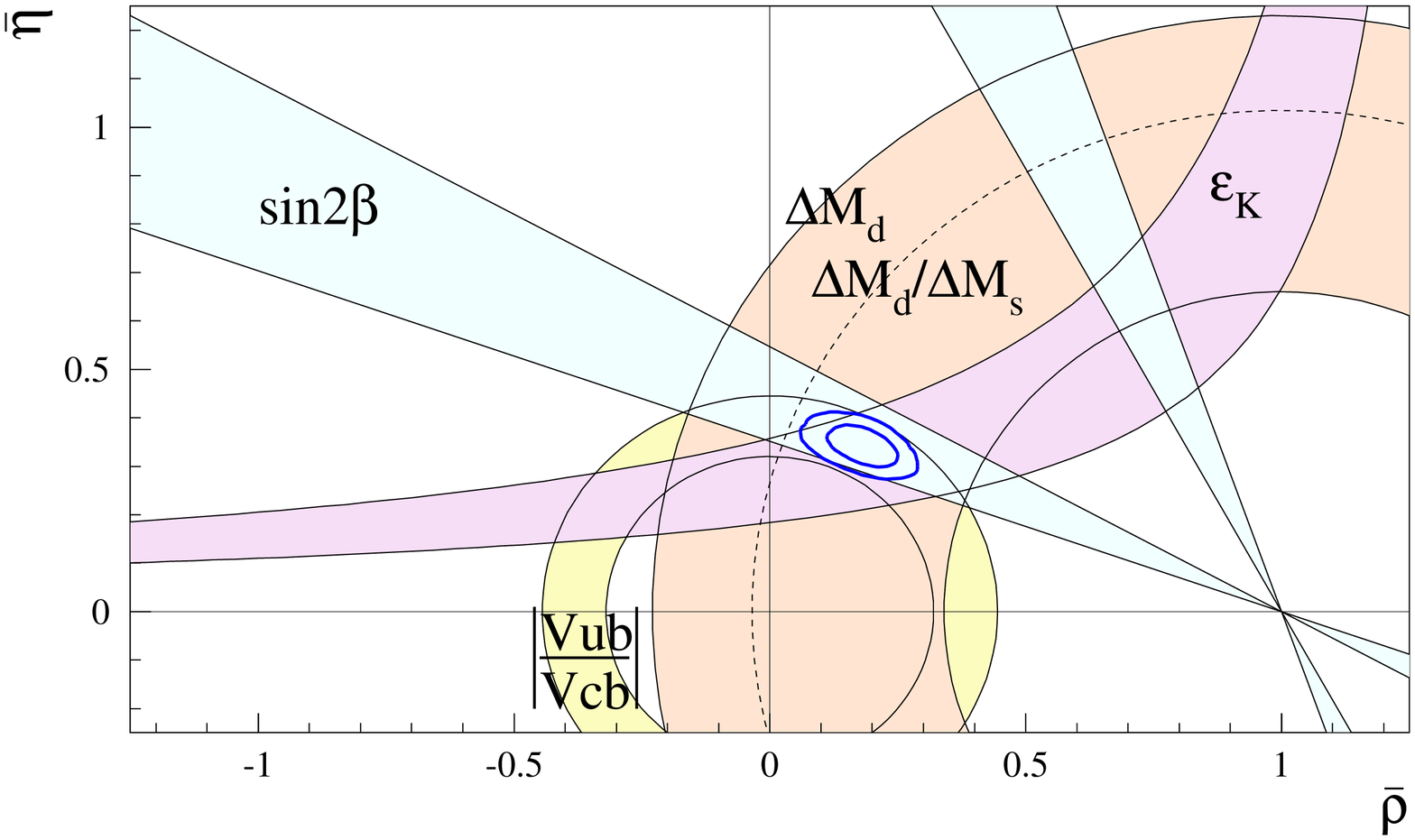}
\hss}
\caption{ \it {Allowed regions for $\rhobar$ and $\etabar$ using the para\-meters
listed in Table~\ref{tab:inputs}. The contours at 68\% and 95\% probability are
shown. The full lines correspond to the 95\% probability constraints given by 
the measurements of $\left | V_{ub} \right |/\left | V_{cb}\right |$, 
$\epsilonk$, $\Delta m_d$ and sin2$\beta$. The dotted curve bounds the region
selected at 95\% by the lower limit on $\Delta m_s$.}}
\label{fig:rhoeta}
\end{figure}

\section{Unitarity Triangle Analysis in the Standard Model}
\label{sec:utasm}

In this section we present the results of the UT analysis assuming the validity
of the SM. These results are obtained by implementing a Bayesian statistical 
analysis~\cite{uta00} using five independent constraints coming from the 
determinations of $\left | V_{ub} \right |/\left | V_{cb}\right |$, $\Delta 
{m_d}$, $\Delta {m_s}/\Delta {m_d} $, $\epsilon_K$ and sin2$\beta$. The regions 
selected by these constraints in the ($\rhobar,~\etabar)$ plane are shown in 
Figure \ref{fig:rhoeta}.

\subsection{Input parameters}
The central values and errors of the input parameters are compared in Table 
\ref{tab:inputs} with those available last year at the time of the first CKM
Workshop. We discuss here the main variations in central values and/or errors 
with respect to last year, referring to ref.~\cite{yb} for supporting details.

On the experimental side, the accuracy reached in the determinations of $\Delta
m_d$, sin2$\beta$ and also of the $b\to u$ semileptonic transitions has further
increased. Concerning the $B^0_s-\bar{B}^0_s$ oscillations, the likelihood now
prefers slightly smaller values of $\Delta m_s$, so that the lower limit on this
quantity is slightly decreased while the sensitivity remains almost unchanged.
In addition, a more critical look at the experimental results leading to the 
determination of the Cabibbo angle $\lambda$ has produced a significant increase
on both the central value and the uncertainty assigned to this 
quantity~\cite{yb}, the latter being increased by approximately 80\% with 
respect to last year. Since, however, $\lambda$ is known with a very good 
relative accuracy, these changes have a marginal impact on the results of the 
UT analysis.

On the theoretical side, a significant improvement concerns the uncertainty on
the inclusive determination of $\vert V_{cb}\vert$ which is reduced, at present,
to the impressive level of 1\%. Another important change, which is worth
mentioning, concerns the theoretical estimate of the so called ``chiral logs
effects" in the lattice evaluation of $\fbdsqbd$, which is mostly reflected in
the final estimate of $\xi$ (see Table \ref{tab:inputs}). For this ratio, the
central value has increased by approximately 5\% but the corresponding relative
uncertainty has been estimated to increase by almost 50\%. Since the estimate 
of this uncertainty is not based directly on lattice data, one can reasonably 
expect that it will be reduced in a rather short time~\cite{damir}.

\subsection{The apex of the UT: $\rhobar$ and $\etabar$}
By using all five constraints ($\left | V_{ub} \right |/\left | V_{cb} \right|$,
 $\Delta {m_d}$, $\Delta {m_s}/\Delta {m_d} $, $\epsilon_K$ and sin2$\beta$), 
the following results for $\rhobar$ and $\etabar$ are obtained
\begin{eqnarray}
\rhobar =  0.178 \pm 0.046 && [0.085-0.265] {\rm ~at ~95\% ~C.L.} \nonumber \\
\etabar =  0.341 \pm 0.028 && [0.288-0.397] {\rm ~at ~95\% ~C.L.}
\end{eqnarray}
Figure \ref{fig:rhoeta} shows the region in the ($\rhobar,~\etabar)$ plane 
selected by the contours at 68\% and 95\% probability.

\subsection{The angles: sin2$\beta$, sin2$\alpha$ and $\gamma$.}

\begin{figure}[htb!]
\hbox to\hsize{\hss
\includegraphics[width=0.4\hsize]{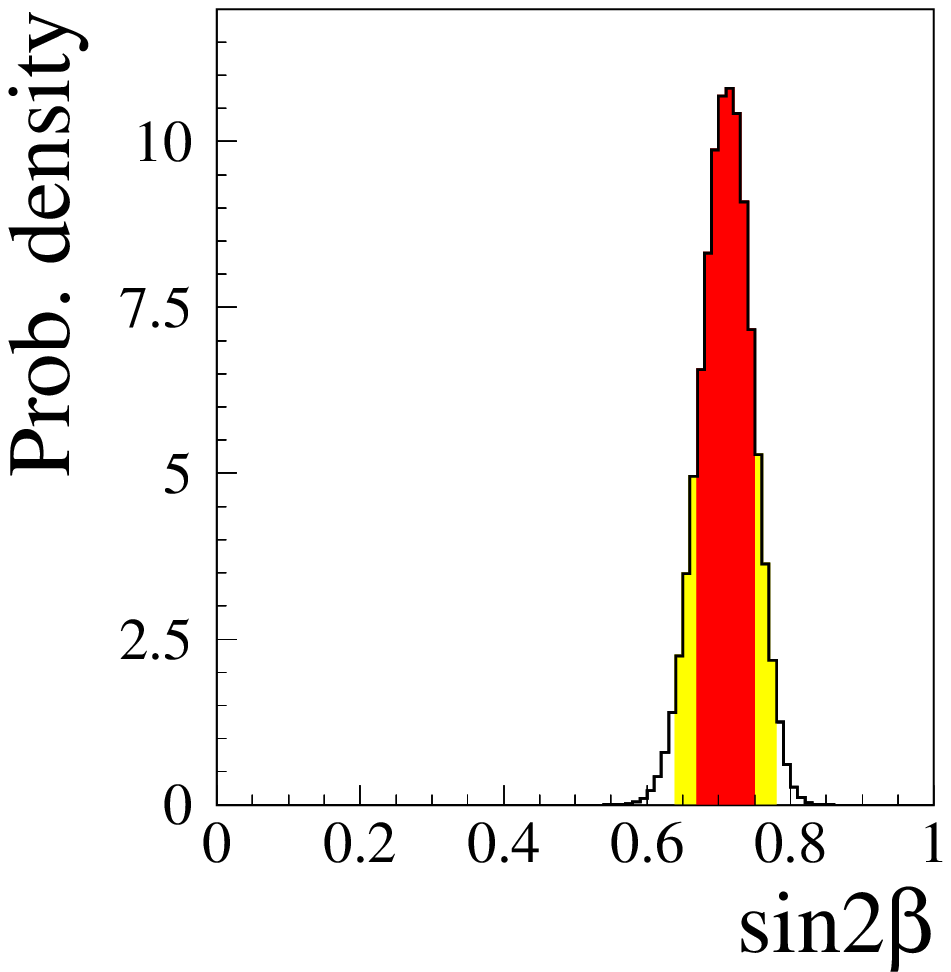}
\hss
\includegraphics[width=0.4\hsize]{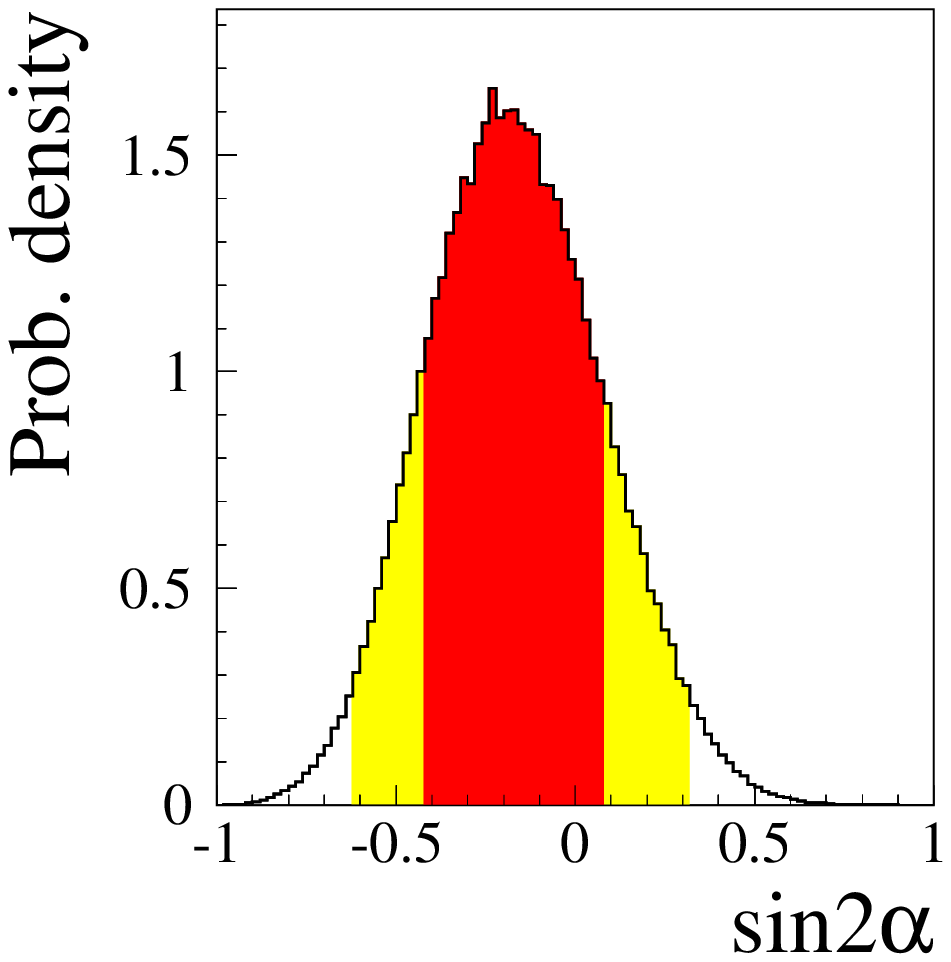}
\hss}
\hbox to\hsize{\hss
\includegraphics[width=0.4\hsize]{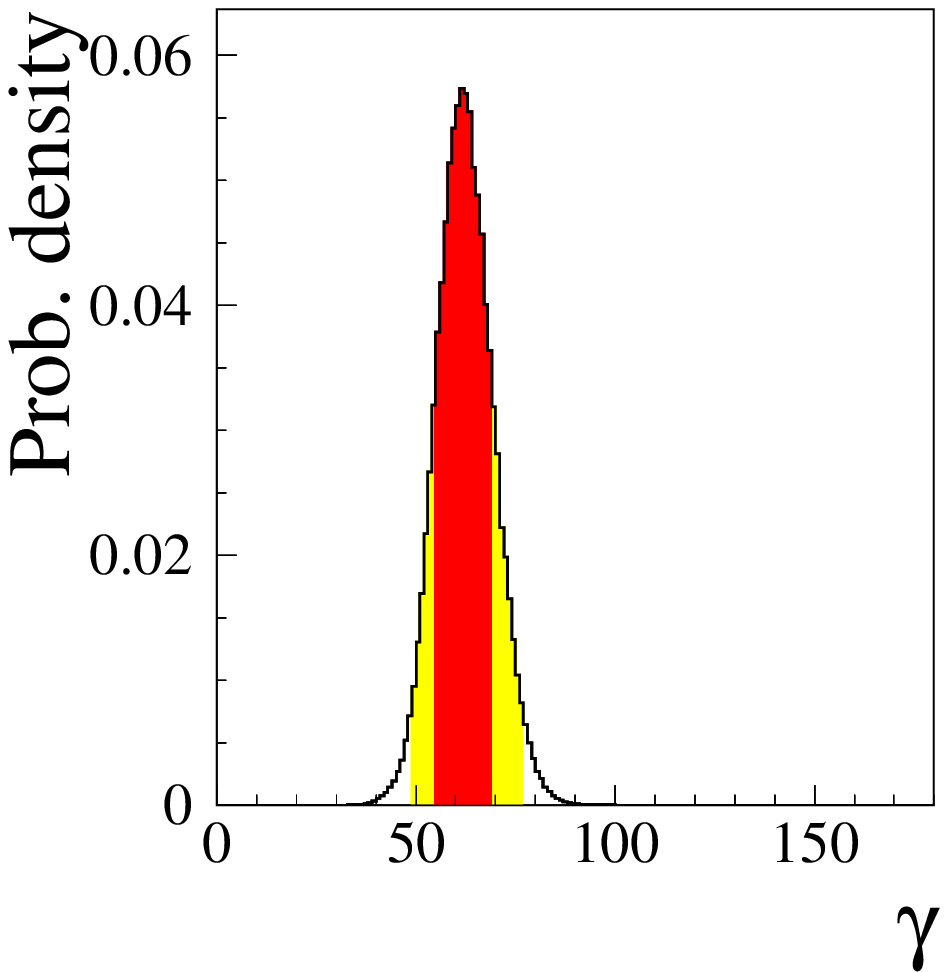}
\hss}
\caption{\it The p.d.f. for sin2$\beta$, sin2$\alpha$ and $\gamma$.}
\label{fig:angles}
\end{figure}

The p.d.f. obtained for sin2$\beta$, sin2$\alpha$ and $\gamma$ are shown in 
Figure \ref{fig:angles}.

It is useful to recall that one of the most important results of the UT analysis
was the prediction of sin2$\beta$. The value of sin2$\beta$ was determined, well
before its first direct measurement, by using all the other available 
constraints, namely $\left | V_{ub} \right |/\left | V_{cb} \right |$, 
$\epsilonk$, $\Delta m_d$ and $\Delta m_s$. The accuracy of these predictions 
was rather good. For instance, at the end of 2000 the following indirect
determination was obtained~\cite{uta00}
\begin{equation}
\sin 2\beta = 0.698 \pm 0.066 \quad \rm {all~constraints~but~A_{CP}}
\label{eq:sin2beta00}
\end{equation}
to be compared with the present direct measurement
\begin{equation}
\sin 2\beta = 0.734 \pm 0.054 \quad A_{CP}(J/\psi K_s). 
\label{eq:sin2beta_dir}
\end{equation}

The indirect determination obtained today, by using the values of the input
parameters collected in Table \ref{tab:inputs}, is
\begin{eqnarray}
\sin 2\beta = 0.685 \pm 0.052  \quad \rm {all~constraints~but~A_{CP}}
\label{eq:sin2beta_today}
\end{eqnarray}
It can be noticed that the present precision for the direct and the indirect 
measurements are similar. When combined, one obtains the best estimate
\begin{equation}
\sin 2 \beta = 0.705 ^{+0.042}_{-0.032} \quad [0.636-0.779] {\rm ~at~95\% ~C.L.}
\label{eq:sin2beta_all}
\end{equation}

The p.d.f. of sin2$\alpha$, presented in Figure \ref{fig:angles}, shows that 
the angle $\alpha$ is much less constrained by the UT analysis than the angle
$\beta$. This distribution corresponds to the result
\begin{equation}
\sin 2 \alpha = -0.19 \pm 0.25 \quad [-0.62-0.33] {\rm ~at ~95\% ~C.L.}
\label{eq:sin2alpha}
\end{equation}

The angle $\gamma$ is predicted with an accuracy which is at present of about 
10\%:
\begin{equation}
\gamma =  (61.5 \pm 7.0)^{\circ} \quad [49.0-77.0]^{\circ}{\rm ~at ~95\% ~C.L.}
\label{eq:gamma}
\end{equation}
The effects of chiral logs in the lattice evaluation of $\xi$ are included in this
determination, although the change in $\gamma$ is not significant (the previous
value was $59.5^{+6.5}_{-5.5}$~\cite{parodi}). It should also be stressed that,
with present measurements, the probability that $\gamma$ is greater than
90$^{\circ}$ is only 0.003.

\subsection{Indirect evidence of CP violation}

\begin{figure}
\hbox to\hsize{\hss
\includegraphics[width=\hsize]{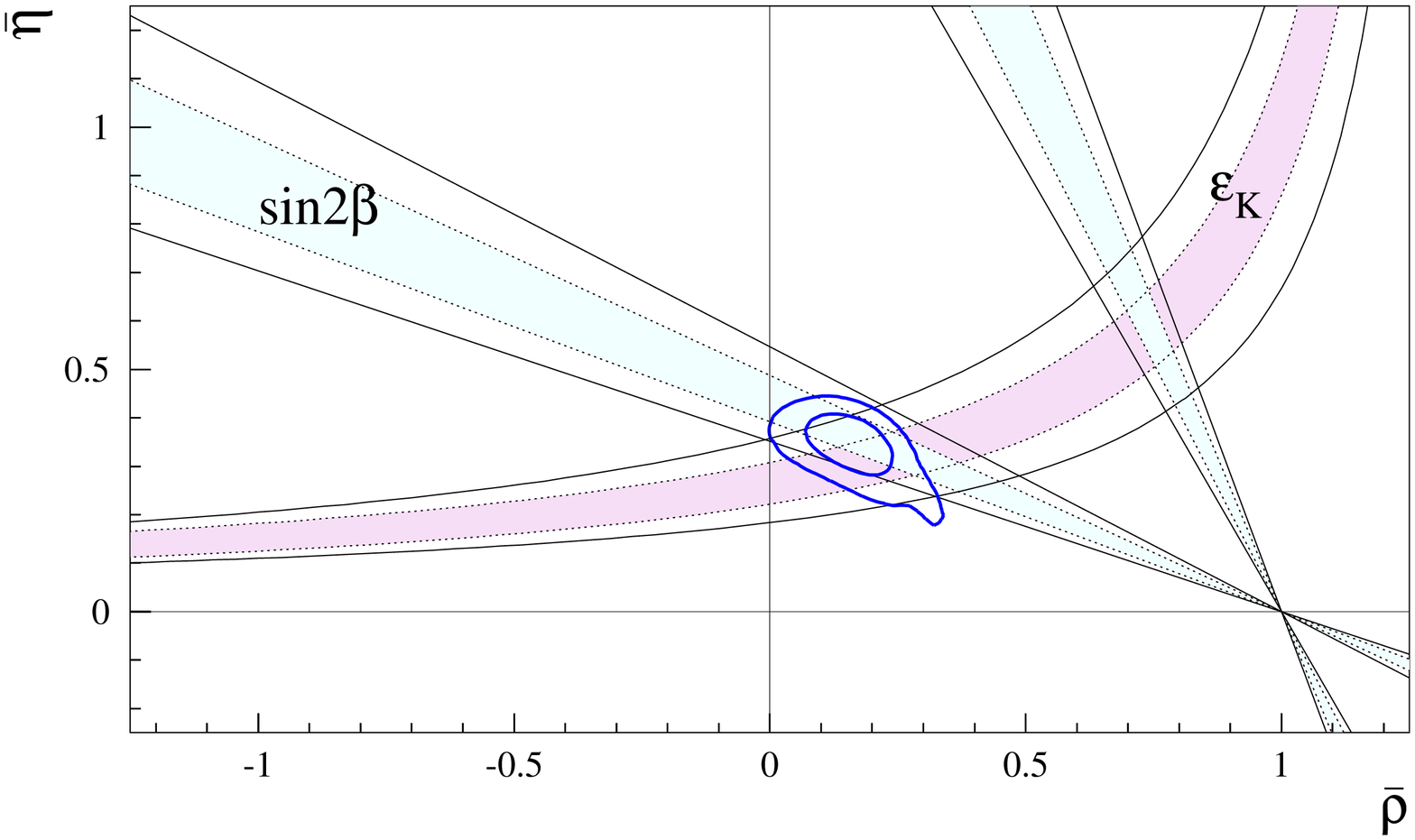}
\hss}
\caption{\it The allowed regions for $\overline{\rho}$ and $\overline{\eta}$
(contours at 68\%, 95\%) as selected by the measurements of $\left | V_{ub} 
\right |/\left | V_{cb} \right |$, $\Delta {M_d}$, and by the limit on $\Delta 
{M_s}/\Delta {M_d} $ are compared with the bands (at 68\% and 95\% C.L.) from 
the measurements of CP violating quantities in the kaon ($\epsilon_K$) and in 
the B (sin2$\beta$) sectors.}
\label{fig:testcp}
\end{figure}

An important test of the SM in the UT analysis is the comparison between the 
region selected by the measurements which are sensitive only to the sides of 
the UT (CP conserving semileptonic B decays and $B^0-\bar{B^0}$ oscillations) 
and the regions selected by the direct measurements of CP violation in the kaon 
($\epsilon_K$) and in the B (sin2$\beta$) sectors. This test is shown in Figure 
\ref{fig:testcp}. 

The result can be made quantitative by comparing the values of sin2$\beta$ 
obtained from the measurement of the CP asymmetry in the $J/\psi K_s$ decays 
with those determined from ``sides" measurements only:
\begin{eqnarray}
\sin 2 \beta = 0.695 \pm 0.056 & \rm {Sides~ only} \nonumber \\
\sin 2 \beta = 0.734 \pm 0.054 & \rm A_{CP}(J/\psi K_s). 
\label{eq:sin2beta}
\end{eqnarray}
The spectacular agreement between these results illustrates the consistency of 
the SM in describing the CP violation phenomena through the CKM mechanism, in 
terms of a single parameter $\eta$. Moreover, it provides an important test of 
the calculations based on the OPE, the HQET and the lattice QCD approaches 
which have been used to extract the CKM parameters.

\subsection{Determination of  $\dms$}
\begin{figure}
\hbox to\hsize{\hss
\includegraphics[width=0.4\hsize]{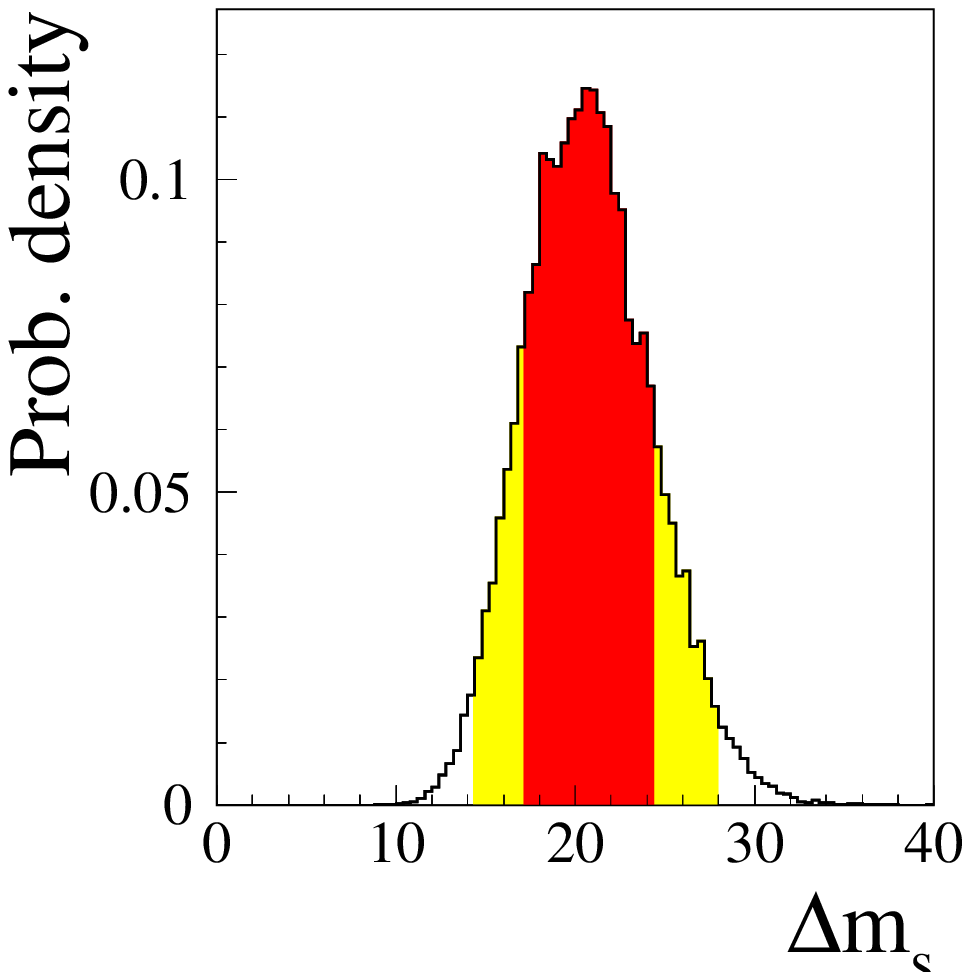}
\hss
\includegraphics[width=0.4\hsize]{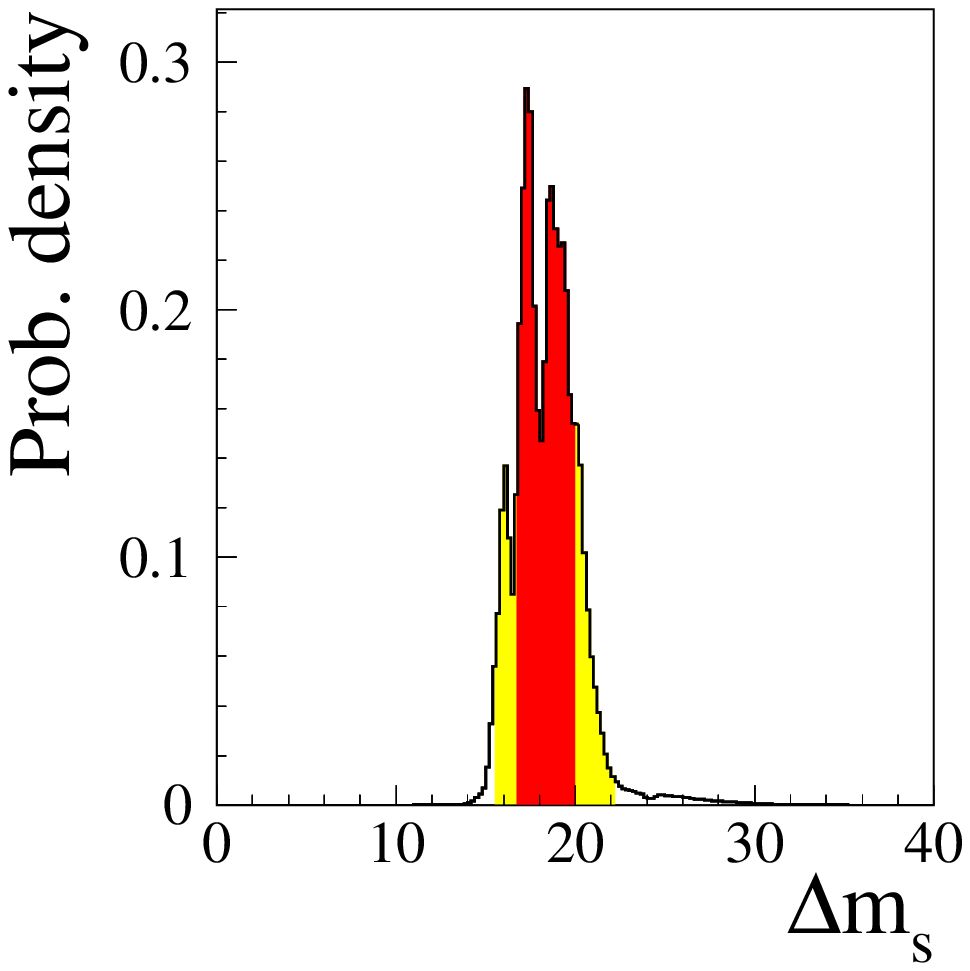}
\hss}
\caption{ \it {$\Delta m_s$ probability distributions. The information from 
${B}^0_s-\bar{{B}}^0_s$ oscillations is used (not used) in the right (left).}}
\label{fig:dms}
\end{figure}

Another important result of the UT analysis is the possibility to extract the 
probability distribution for the mass difference $\dms$, which is shown in 
Figure \ref{fig:dms}. 
The corresponding results, obtained either by using or not using the 
experimental information coming from the analysis of the ${B}^0_s-\bar{{B}}^0_s$
oscillations, are
\begin{eqnarray}
\label{eq:dms}
\dms = 18.3^{+1.7}_{-1.5} & [15.6-22.2]{\rm ~at ~95\% ~C.L.} \nonumber \\
& ~~~~~~~~~~~~~~~~~~~~\rm{with ~\dms} \nonumber \\
\dms = 20.6 \pm 3.5 & [14.2-28.1]{\rm ~at ~95\% ~C.L.} \nonumber \\
& ~~~~~~~~~~~~~~~~~\rm{without ~\dms}
\end{eqnarray}

The present limit (from LEP and SLD) excludes already a fraction of the $\dms$ 
distribution. Present analyses at LEP/SLD are situated in a high probability 
region for a positive signal (as the ``signal bump'' appearing around 17 
ps$^{-1}$). Accurate measurements of $\dms$ are expected soon from the TeVatron.
The result in the second of Equations~(\ref{eq:dms}) represents another
significant prediction of the UT analysis.

\subsection{Determination of $\fbdsqbd$ and $B_K$}
Since the UT fits in the SM are currently overconstrained, they can also be used
to extract a determination of one (or some) of the relevant hadronic parameters 
entering the analysis. This determination then provides a useful comparison for 
the corresponding theoretical predictions obtained from lattice QCD simulations.

\begin{figure}
\hbox to\hsize{\hss
\includegraphics[width=0.4\hsize]{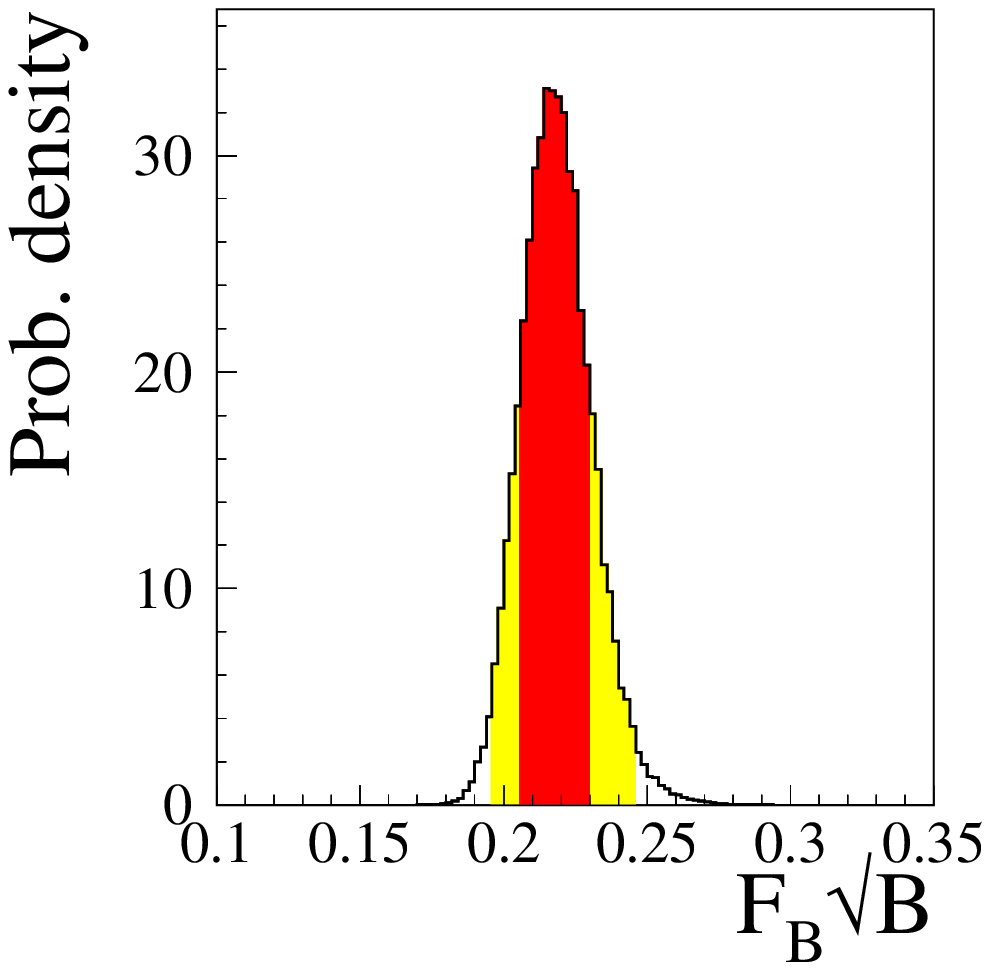}
\hss
\includegraphics[width=0.4\hsize]{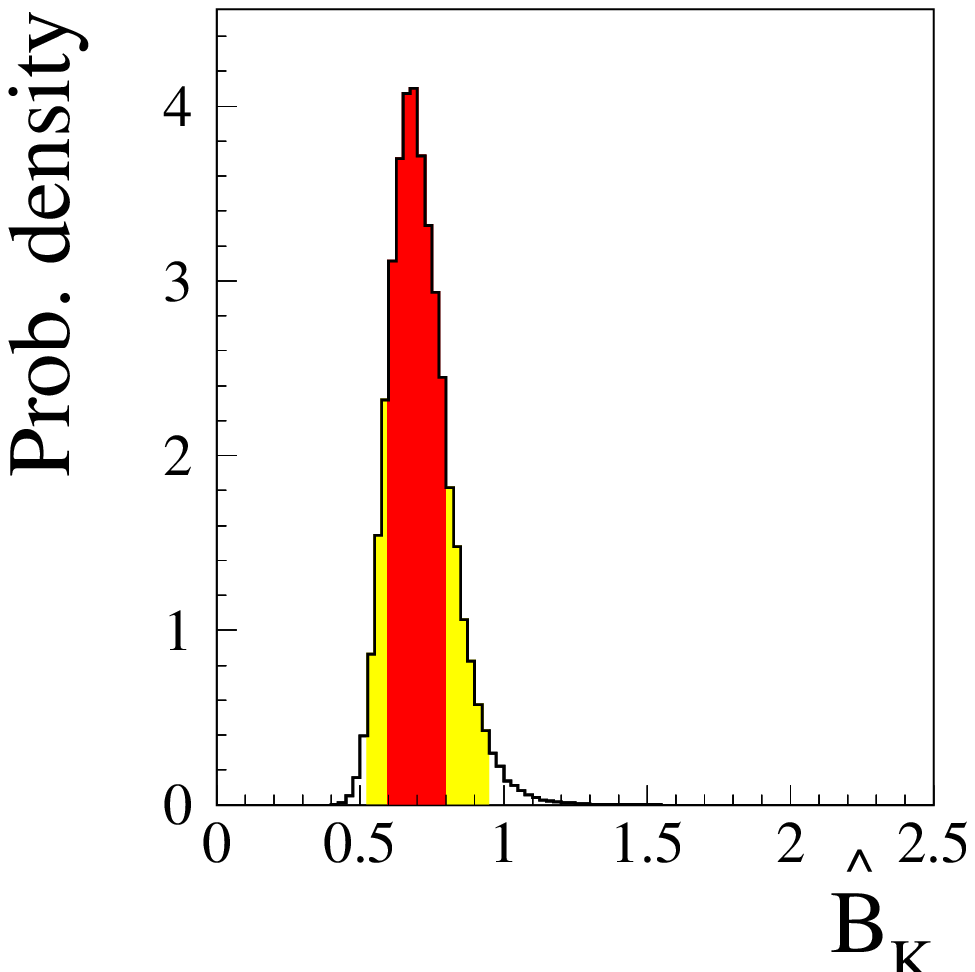}
\hss}
\caption{ \it {The p.d.f. for $\fbdsqbd$ (left) and $B_K$ (right).}}
\label{fig:fbbk}
\end{figure}

The p.d.f. of $\fbdsqbd$ is shown in Figure~\ref{fig:fbbk} (left).
From this distribution one obtains
\begin{equation}
\fbdsqbd = (217 \pm 12) \, \MeV  \,
\end{equation}
which is in very good agreement with the lattice determination given in Table 
\ref{tab:inputs}. Notice also that $\fbdsqbd$ is determined with an accuracy 
which is better than the current evaluation from lattice QCD. 

The density distribution for the parameter $B_K$ is given in Figure 
\ref{fig:fbbk} (right). The fitted value is
\begin{equation}
B_K = 0.69^{+0.11}_{-0.09}\, ,
\label{eq:bk}
\end{equation}
which is in agreement, though one standard deviation smaller, with the 
corresponding lattice estimate. By looking at the uncertainty obtained in 
Equation (\ref{eq:bk}) one concludes that the present estimate of $B_K$ from
lattice QCD, with a 15\% relative error, has a large impact in the present 
analysis.

Notice that values of $B_K$ smaller than 0.5 (0.3) have a probability of 0.6\% 
(5 $\times$ 10$^{-6}$), whereas values of $B_K$ larger than 1.0 have a 
probability of 1.3\%. In disfavoring large values of $B_K$ the direct 
measurement of sin2$\beta$ plays a crucial role.

The results above on either $\fbdsqbd$ or $\hat \BK$ are obtained by using the 
theoretical information coming from the distributions of both the other 
parameter and $\xi$. It is interesting to see if significant results can be 
obtained by removing simultaneously two of the previous constraints.

\begin{figure}
\hbox to\hsize{\hss
\includegraphics[width=\hsize]{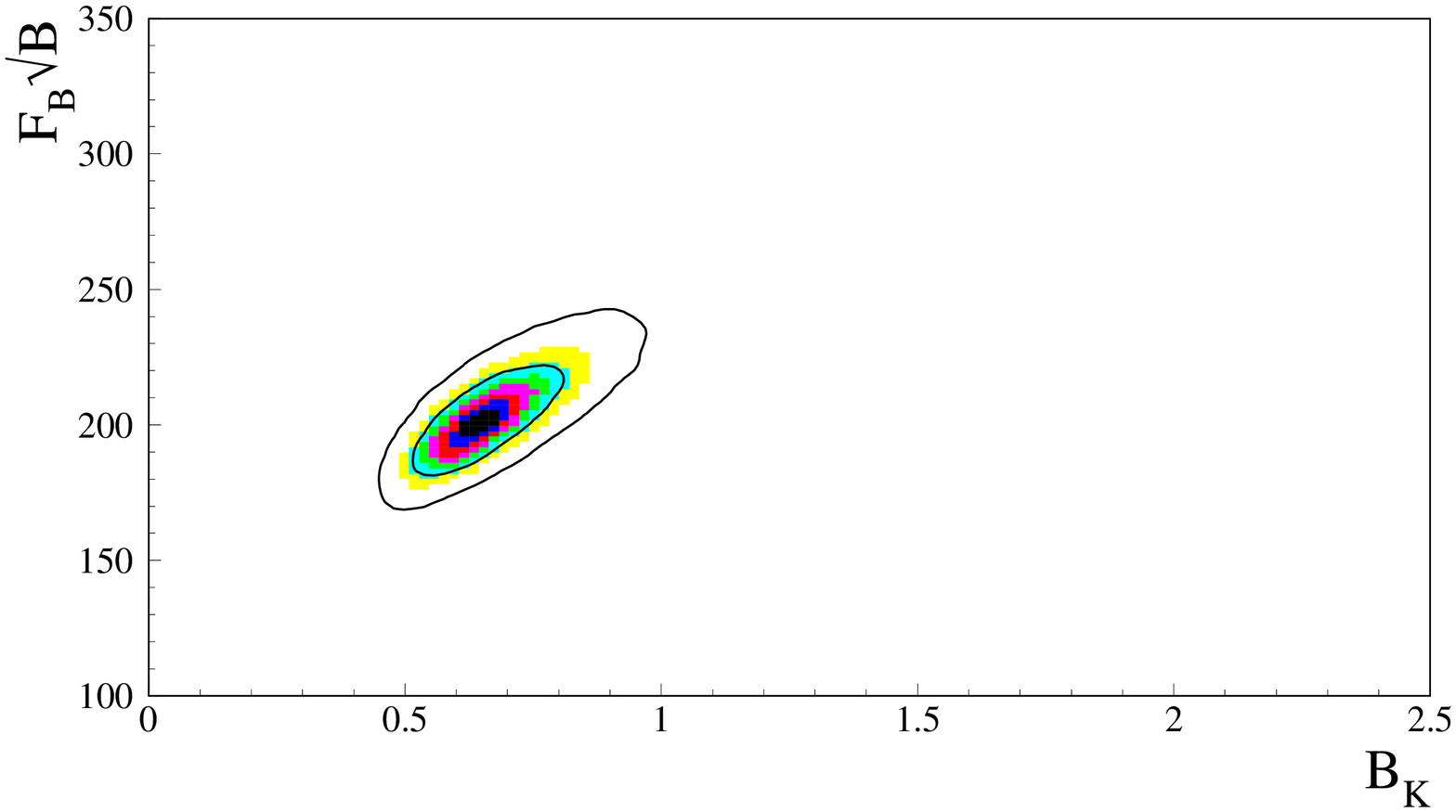}
\hss}
\hbox to\hsize{\hss
\includegraphics[width=\hsize]{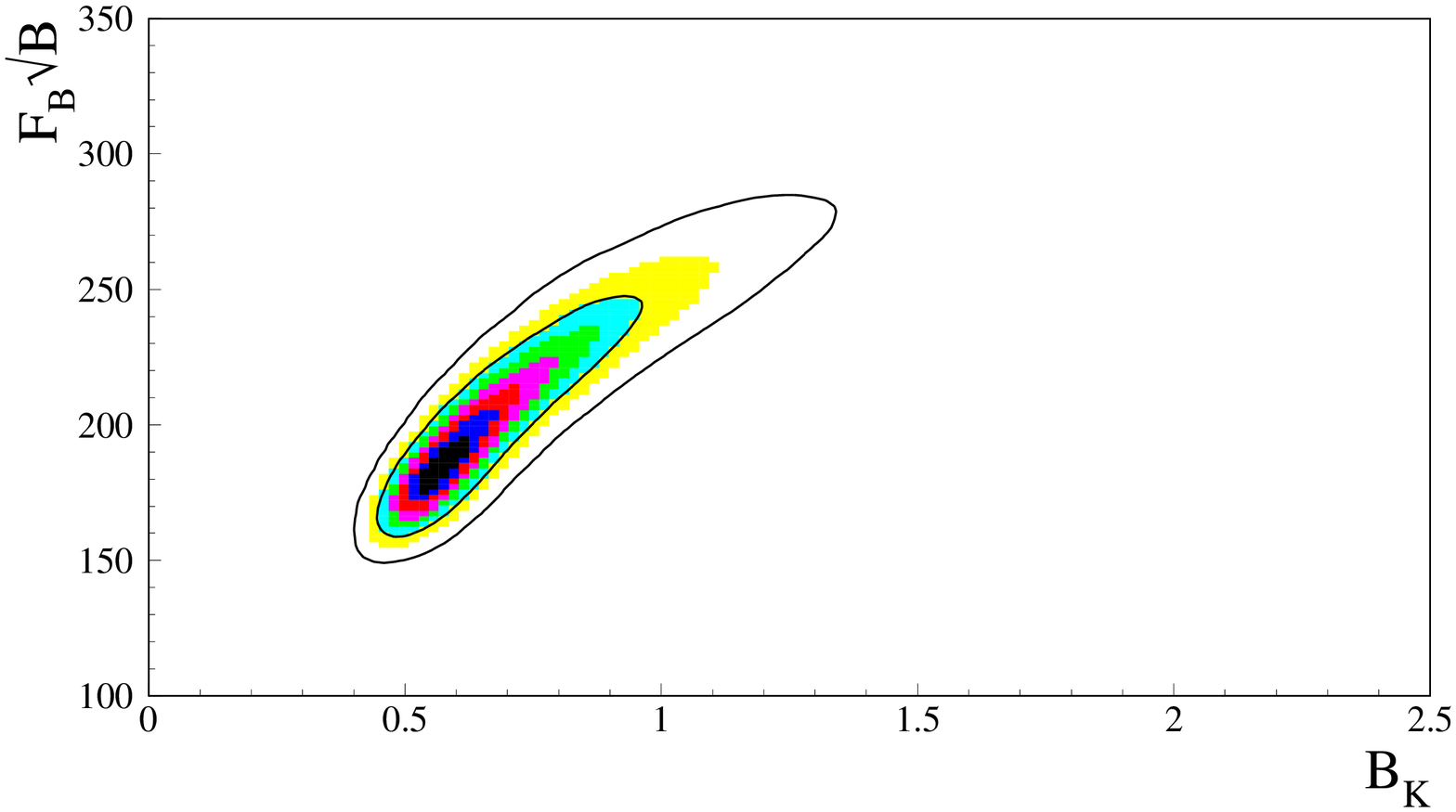}
\hss}
\caption{\it { The 68\% and 95\% contours in the $(\fbdsqbd$,
$\hat B_K)$ plane. The plot on the top is obtained using the $\left | V_{ub} 
\right |/\left | V_{cb} \right |$, $\Delta m_s$ and sin2$\beta$ constraints, 
while in the bottom one also the information from $\Delta m_s$ (and thus $\xi$)
has been removed.}}
\label{fig:bkfb}
\end{figure}

The  region in the plane ($\fbdsqbd$,~$\hat B_K$), which is 
obtained by removing the theoretical constraints on these quantities, is shown 
in Figure~\ref{fig:bkfb} (top). Within 68\% and 95\% probabilities, both 
$\fbdsqbd$ and $\hat B_K$ are well constrained. 
The most important conclusion which can be drawn from this study is the 
simultaneous range obtained for $\fbdsqbd$ and $\hat B_K$, namely
\begin{eqnarray}
& \hat B_K = 0.69^{+0.13}_{-0.08} &  [0.53-0.96] \\ \nonumber
& \fbdsqbd = 203^{+17}_{-13}~{\rm MeV} &  [180-242]~{\rm MeV} 
\end{eqnarray}
The same analysis can be performed by removing also the information coming from
$\dms$, namely removing the external constraint controlled by the non 
perturbative QCD parameter $\xi$. The region obtained in the plane ($\fbdsqbd$, 
$\hat B_K$) is shown in Figure \ref{fig:bkfb} (bottom) and corresponds to
\begin{eqnarray}
& \hat B_K = 0.67^{+0.26}_{-0.13} &  [0.47-1.27] \\ \nonumber
& \fbdsqbd = 203^{+38}_{-28}~{\rm MeV} &  [162-278]~{\rm MeV} 
\end{eqnarray}

This last result relies only on the $\left| V_{ub}\right|/\left| V_{cb}\right|$
and sin2$\beta$ constraints, and it is thus only dependent on the hadronic 
parameters entering the $B$ semileptonic decays.

\section{New Physics: impact of future (and more precise) measurements}
\label{sec:np1}

In this section we would like to discuss the interest of measuring the various 
physical quantities entering the UT analysis with a better precision. We 
investigate, in particular, to which extent future and improved determinations 
of the experimental constraints, sin2$\beta$, $\dms$ and $\gamma$, could allow 
us to possibly invalidate the SM, thus signalling the presence of NP effects.

\subsection{Sin2$\beta$}
\begin{figure}
\hbox to\hsize{\hss
\includegraphics[width=0.7\hsize]{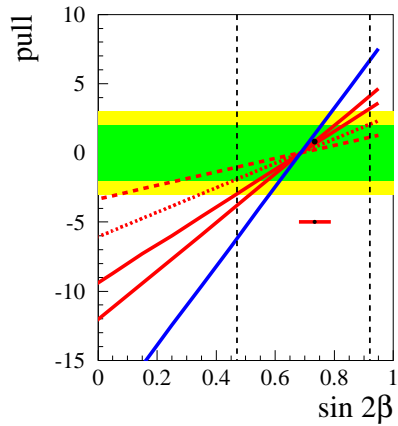}
\hss}
\caption{\it {The compatibility (``pull'') between the direct and indirect 
determination of sin2$\beta$ as a function of the value of sin2$\beta$ measured 
from CP asymmetry in $J/\psi K_s$ decays. The different curves correspond to 
different errors for sin2$\beta$ (from top to bottom 0.2, 0.1, 0.05 and 0.02). 
The last curve (darker) corresponds to the scenario in which all the 
experimental and theoretical errors contributing to the indirect measurement of 
sin2 $\beta$ are reduced by a factor of two. The point indicate the situation 
for the present measurement. The horizontal bands indicate the 2 and 3 $\sigma$ 
compatibility regions. The 3$\sigma$ region which corresponds to an error of 
0.05 on sin2$\beta$ is indicated by the vertical dotted lines.}}
\label{fig:pull_sin2b_senza}
\end{figure}

We start this analysis by considering the measurement of sin2$\beta$. The plot 
in Figure \ref{fig:pull_sin2b_senza} shows the compatibility (``pull'') between 
the direct and indirect distributions of sin2$\beta$, in the SM, as a function 
of the (hypothetical) measured value of sin2$\beta$, parametrized for different 
values of the errors. The compatibility between the two distributions is 
determined by using the p.d.f. of the difference between the direct and the 
indirect determinations. The compatibility is then evaluated by mean of the 
ratio between the central value of this resulting (pseudo-Gaussian) distribution
and its standard deviation. 

From the plot in Figure \ref{fig:pull_sin2b_senza} it can be seen that, 
considering the actual precision of about 0.05 on the measured value of 
sin2$\beta$, the 3$\sigma$ compatibility region is between [0.46-0.92]. Values 
outside this range would be, therefore, not compatible with the SM prediction at
more than 3$\sigma$ level. To get these values, however, the presently measured 
central value should shift by more than 4$\sigma$.

Figure \ref{fig:pull_sin2b_senza} also shows that, in case the experimental 
error on sin2$\beta$ was 0.02, the compatibility region would be reduced to 
[0.52-0.86]. Furthermore, if all the errors contributing to the indirect
determination of sin2$\beta$ are decreased by a factor of two that region is 
reduced by more than a factor 1.5 (as shown by the darker curve).

The conclusion that can be derived from Figure \ref{fig:pull_sin2b_senza} is
the following: although the improvement of the error sin2$\beta$ has an 
important impact on the accuracy of the UT parameter determination, it is very 
unlikely that in the near future we will be able to use this measurement to 
detect any failure of the SM, unless the central value of the direct measurement
will move away from the present one by several standard deviations.

It was pointed out sometime ago that the comparison of the time dependent CP 
asymmetries in various $B$-decay modes could provide evidence of 
NP~\cite{ref:phik01}. Beside the $J/\psi K_s$ mode, the asymmetry $A_{CP}(\phi 
K_s)$ also allows to extract sin2$\beta$ with negligible hadronic uncertainties 
\cite{ref:phik02}. Furthermore the $\phi K_s$ mode is expected to be more 
sensitive to NP being a pure penguin process.

This asymmetry has been recently measured at the B-factories 
\cite{ref:phik0_mea}:
\begin{eqnarray}
\sin 2 \beta =  -0.39 \pm 0.41 & \quad \phi K_s~ {\rm mode}
\label{eq:sin2beta_phik0}
\end{eqnarray}

\begin{figure}
\hbox to\hsize{\hss
\includegraphics[width=0.7\hsize]{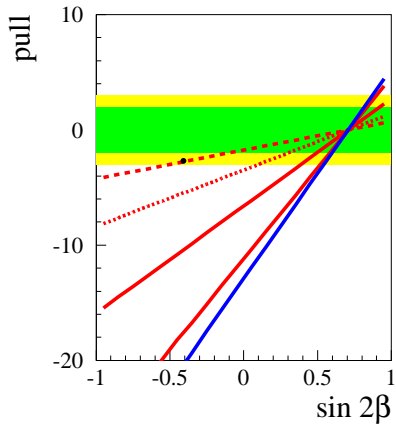}
\hss}
\caption{\it {The compatibility of the direct and indirect determination of
sin2$\beta$ as a function of the value of sin2$\beta$ measured from CP 
asymmetry in $\phi K_s$ decays. The different curves correspond to 
different errors for sin2$\beta$ (from top to bottom 0.4, 0.2, 0.1 and 0.05). 
The point indicate the situation for the present measurement. The last curve 
(darker) corresponds to the scenario in which all the experimental and 
theoretical errors contributing to the indirect measurement of sin2 $\beta$ are 
reduced by a factor of two. The horizontal bands indicate the 2 and 3 $\sigma$ 
compatibility regions.}}
\label{fig:pull_sin2b_con}
\end{figure}
The plot in Figure \ref{fig:pull_sin2b_con} shows the compatibility of the 
direct and indirect distributions of sin2$\beta$ as a function of the measured 
value of sin2$\beta$ parametrized for different errors. The difference with
respect to the plot in Figure \ref{fig:pull_sin2b_senza} is that in this case 
all the available constraints have been used to obtain the indirect distribution
of sin2$\beta$, including the direct measurement of sin2$\beta$ from 
$J/\psi K_s$.

It can be noticed that the current measured value in the $\phi K_s$ mode is at 
2.7$\sigma$ deviation from the expected one. This deviation is obviously
dominated by the experimental error on the measured CP asymmetry, and its 
significance is not really affected if some or even all the theoretical errors 
are multiplied by a factor of two. Keeping the same central value a reduction of
a factor of two in the experimental error will shift the significance to 
5.7$\sigma$ and any measured value below zero will deviate by more than 
4$\sigma$ from the SM prediction.

\subsection{$\dms$}
\begin{figure}
\hbox to\hsize{\hss
\includegraphics[width=0.7\hsize]{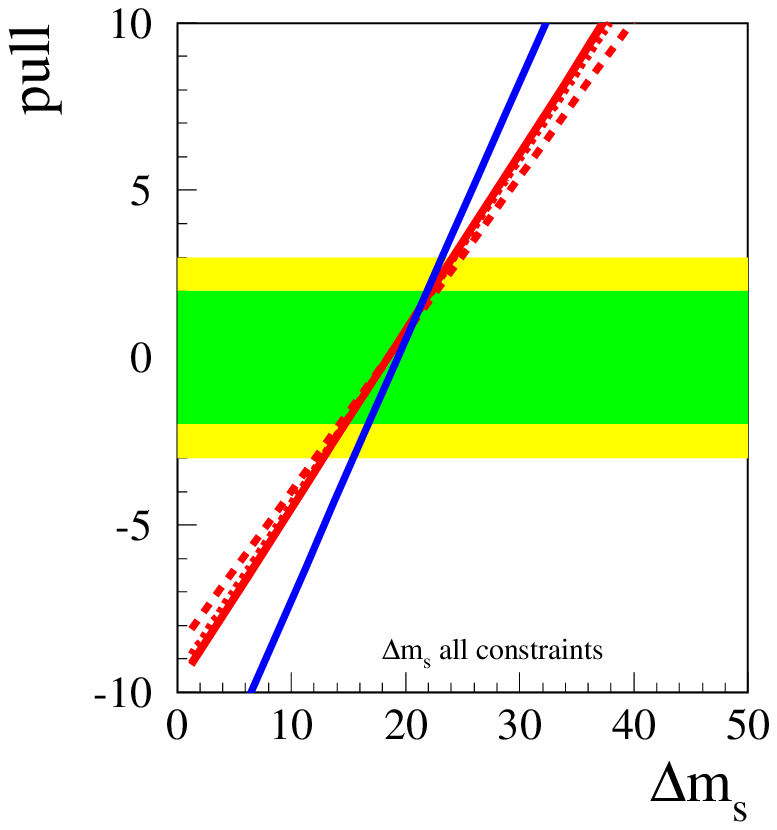}
\hss}
\caption{\it {Same as in Figure \ref{fig:pull_sin2b_con} but for $\dms$. The 
different curves correspond to different errors for this determination (from 
bottom to top 1, 0.5, 0.2 and 0.1 $ps^{-1}$).}}
\label{fig:pull_dms}
\end{figure}

The plot in Figure \ref{fig:pull_dms} shows the compatibility of the indirect 
determination of $\dms$ with a future determination of the same quantity.
It can be noted that the predicted ranges for $\dms$ which are compatible within
3$\sigma$ with the SM prediction do not really depend on the accuracy of the 
future measurement. Furthermore these ranges will be not much affected by a 
reduction of the errors of the quantities entering in the indirect 
determination. From the plot in Figure \ref{fig:pull_dms} it can be concluded
that 
\begin{eqnarray}
\dms > 29\,(28)\,[26]~{\rm ps}^{-1}  & {\rm ~``New~ Physics''~ at~ 5}~\sigma \\ 
\nonumber
\dms >  25\,(24)\,[23]~{\rm ps}^{-1}  &  {\rm ~``New~ Physics''~ at~ 3}~\sigma  
\end{eqnarray}
corresponding to 1~ps$^{-1}$ (0.1~ps$^{-1}$) [0.1~ps$^{-1}$ with all the 
errors divided by two] scenarios.

\subsection{The angle $\gamma$}

\begin{table*}[t]
\begin{center}
\begin{tabular}{|c|c|c|c|}
\hline
Tree-Level  & B$^0_d$ mixing (1,3) family & B$^0_s$ mixing (2,3) family & 
K$^0$ mixing (1,2) family  \\ 
\hline
$\left | V_{ub} \right |/\left | V_{cb} \right |$ & $\Delta {m_d}$ and 
$A_{CP}(J/\Psi \phi)$ & $\Delta {m_s}$ & $\epsilon_K$  \\
\hline
\end{tabular}
\caption{ \it Different processes and corresponding measurements contributing 
to the determination of $\rhobar$ and $\etabar$.}
\label{tab:schema}
\end{center} 
\end{table*}
\begin{figure}[t]
\hbox to\hsize{\hss
\includegraphics[width=0.7\hsize]{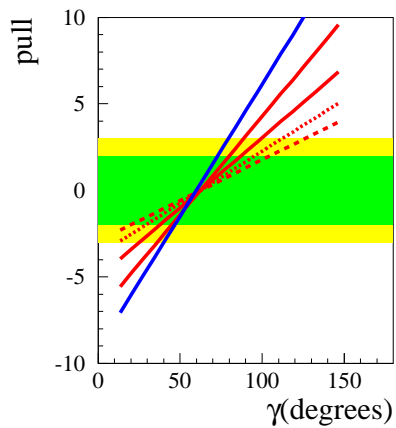}
\hss}
\caption{\it {Same as in Figure \ref{fig:pull_sin2b_con} but for $\gamma$. The 
different curves correspond to different errors for this determination (from 
bottom to top 20, 15, 10 and 5 degrees).}}
\label{fig:pull_gamma}
\end{figure}
The angle $\gamma$ can be also extracted, in principle, from the measurements of
two-body charmless hadronic B decays. A lot of theoretical investigations have 
been recently made for these decay channels. In particular, important progress 
has been made with the calculation of the amplitudes in the heavy quark limit 
using the factorization approach~\cite{bbns}, though there is still some 
controversy on the importance of non-leading corrections~\cite{roma}. 

From the experimental point of view an impressive effort has been made to 
measure as many branching fractions and CP asymmetries as possible (for a 
collection of results see \cite{ref:ichep2002}). It would be very interesting 
in the near future to compare the determination of the angle $\gamma$ from the
UT analysis and from two-body decays measurements. In the future, at the
LHC/BTeV experiments, $\gamma$ will be cleanly measured in the tree-level 
$B_s\to DK$ decays.

The plot in Figure \ref{fig:pull_gamma} shows the compatibility of the indirect 
determination of $\gamma$ with a future determination of the same angle obtained
from B decays. It can be noted that even in case the angle $\gamma$ can be 
measured with a precision of 10$^{\circ}$ from B decays, the predicted 3$\sigma$
region is still rather large, corresponding to the interval [25-100]$^{\circ}$. 
If all the theoretical errors are divided by a factor of two, as indicated by 
the darker curve, the predicted 3$\sigma$ region will be consistently reduced.

\section{New Physics: a (simplified) Model Independent analysis}
\label{sec:np2}
Although the CKM mechanism is extremely successful, leading in particular to 
the precise prediction of the value of sin2$\beta$, it is nevertheless worth to 
investigate whether the analysis of the UT still allows some room for NP 
effects. This is the issue we would like to address in this section.

The physical processes entering the analysis and the related physical 
observables determined from the experiments, are listed in 
Table~\ref{tab:schema}. Barring the possibility of significant NP effects in the
determination of the ratio $\left| V_{ub}\right|/\left| V_{cb} \right|$ from 
tree-level processes, we have explored the possible contributions to 
$B^0_q-\bar{B}^0_q$ mixing ($q=d,s$) and $K^0-\bar{K}^0$ mixing. 

NP contributions introduce in general a large number of new parameters: flavour 
changing couplings, short distance coefficients and matrix elements of new local
operators. The specific list and the actual values of these parameters depend on
the details of the NP model. Nevertheless, each of the mixing process listed in 
Table~\ref{tab:schema}, being described by a single amplitude, can be 
effectively parameterized in a completely general way in terms of only two new 
parameters, which we choose to quantify the difference of the amplitude in 
absolute value and phase with respect to the SM one~\cite{ref:cfactors}. Thus, 
for instance, in the case of $B^0_q-\bar{B}^0_q$ mixings we define
\begin{equation}
C_q \, e^{2 i \phi_q} = \frac{\langle B^0_q|H_{eff}^{full}|\bar{B}^0_q\rangle}
{\langle B^0_q|H_{eff}^{SM}|\bar{B}^0_q\rangle}\, \qquad (q=d,s)
\label{eq:paranp}
\end{equation}
where $H_{eff}^{SM}$ includes only the SM box diagrams, while $H_{eff}^{full}$ 
includes also the NP contributions. By definition, in the absence of NP effects,
$C_q=1$ and $\phi_q=0$. The experimental quantities determined from the 
$B^0_q-\bar{B}^0_q$ mixings and listed in Table~\ref{tab:schema} are related to 
their SM counterparts and the NP parameters by the following relations:
\begin{eqnarray}
& \Delta m_d = C_d ~\Delta m_d^{SM} \qquad \quad
(B^0_d-\bar{B}^0_d~\rm{mixing}) \nonumber \\
& A_{CP}(J/\Psi~ K_s) = \sin 2(\beta+\phi_d) \qquad \qquad \qquad
\label{eq:cfactd}
\end{eqnarray}
and
\begin{eqnarray}
& \Delta m_s = C_s ~\Delta m_s^{SM} & \qquad (B^0_s-\bar{B}^0_s~\rm{mixing})\,.
\label{eq:cfacts}
\end{eqnarray}
As far as the $K^0-\bar{K}^0$ mixing is concerned, we find it convenient to 
introduce a single parameter which relates the imaginary part of the amplitude 
to the SM one
\begin{equation}
C_\epsilon = \frac{{\rm Im}[\langle K^0|H_{eff}^{full}|\bar{K}^0\rangle]}
{{\rm Im}[\langle K^0|H_{eff}^{SM}|\bar{K}^0\rangle]}\, .
\label{eq:ceps}
\end{equation}
This definition implies in fact a simple relation for $\epsilonk$,
\begin{eqnarray}
& \epsilonk = C_{\epsilon}~\epsilonk^{SM} & \qquad 
(K^0-\bar{K}^0~\rm{mixing})   
\label{eq:cfactk}
\end{eqnarray}
Thus, all NP effects which may enter the present analysis of the UT are
parameterized in terms of four real coefficients, $C_d$, $\phi_d$, $C_s$ and
$C_{\epsilon}$. 

Due to the limited available number of constraints, we now make the hypothesis 
that NP effects can appear (at least to a large extent) only in one of the three
mixing amplitudes. This is the only restrictive assumption we make in this 
analysis. Clearly, this hypothesis is not valid in several specific models, as
for instance in the Minimal Flavour Violating SUSY models~\cite{mfv} in which NP
effects simultaneously appear in all the mixing amplitudes but with coefficients
which are related one to the other. 

\begin{figure}[t]
\hbox to\hsize{\hss
\includegraphics[width=0.7\hsize]{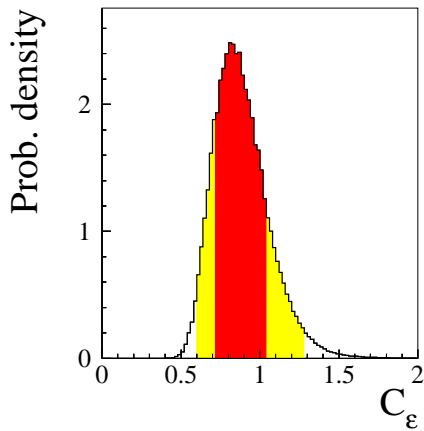}
\hss}
\vspace{-0.2cm}
\caption[]{\it{The $C_{\epsilon}$ distribution obtained by leaving this
parameter free in the fit and assuming the SM parameterization for $\left | 
V_{ub} \right |/\left | V_{cb} \right |$, $\Delta {m_d}$, $\Delta {m_s}/\Delta 
{m_d} $ and sin2$\beta$.}}
\label{fig:cepsk}
\end{figure}

We start the analysis by considering NP contributions in $K^0-\bar{K}^0$ mixing.
Figure \ref{fig:cepsk} shows the distribution of the parameter $C_{\epsilon}$. 
The result from the fit is:
\begin{eqnarray}
C_{\epsilon} = 0.85^{+0.20}_{-0.14}  ~~~~~~   [0.60-1.28]~ \rm{at}~95\% ~C.L.
\label{eq:cepsk}
\end{eqnarray}
The value is compatible with unity but the distribution is rather broad. Thus, 
in this respect, large NP contributions to $K^0-\bar{K}^0$ mixing are still 
allowed. We note, however, that the experimental constraint coming from 
$\epsilonk$ can only determine the product $C_{\epsilon}\cdot B_K$. Therefore, 
the large width of the distribution of $C_{\epsilon}$ simply reflects the
uncertainty existing on the hadronic parameter $B_K$. We also find that, in this
scenario, the distributions of the other UT parameters ($\rhobar$, $\etabar$,
sin2$\beta$, $\ldots$) are not really different from those obtained using the SM
parameterization ($C_{\epsilon}=1$).

\begin{figure}[t]
\hbox to\hsize{\hss
\includegraphics[width=0.7\hsize]{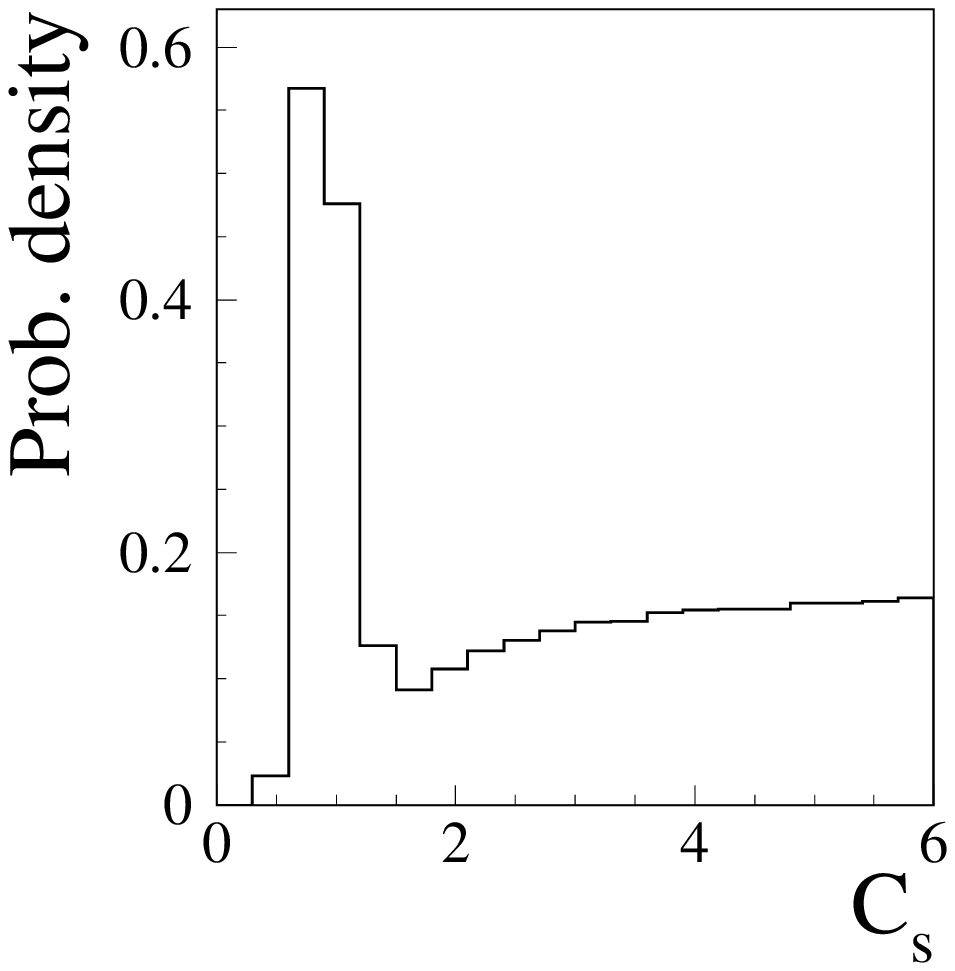}
\hss}
\vspace{-0.2cm}
\caption[]{\it{The $C_s$ distribution obtained by leaving this parameter
free in the fit and assuming the SM parameterization for $\left | V_{ub} \right 
|/\left | V_{cb} \right |$, $\Delta {m_d}$, $\epsilonk$ and sin2$\beta$.}}
\label{fig:cs}
\end{figure}

Figure \ref{fig:cs} shows the distribution of the parameter $C_{s}$ entering the
$B^0_s-\bar{B}^0_s$ mixing amplitude.

The value of $C_{s}$ peaks at 1. However, it is not limited from above due to 
the fact the $\Delta m_s$ is not yet determined. Also in this case, the 
distributions of the other UT parameters are not really different from those 
obtained using the SM parameterization ($C_s=1$).

\begin{figure}[th]
\hbox to\hsize{\hss
\includegraphics[width=0.4\hsize]{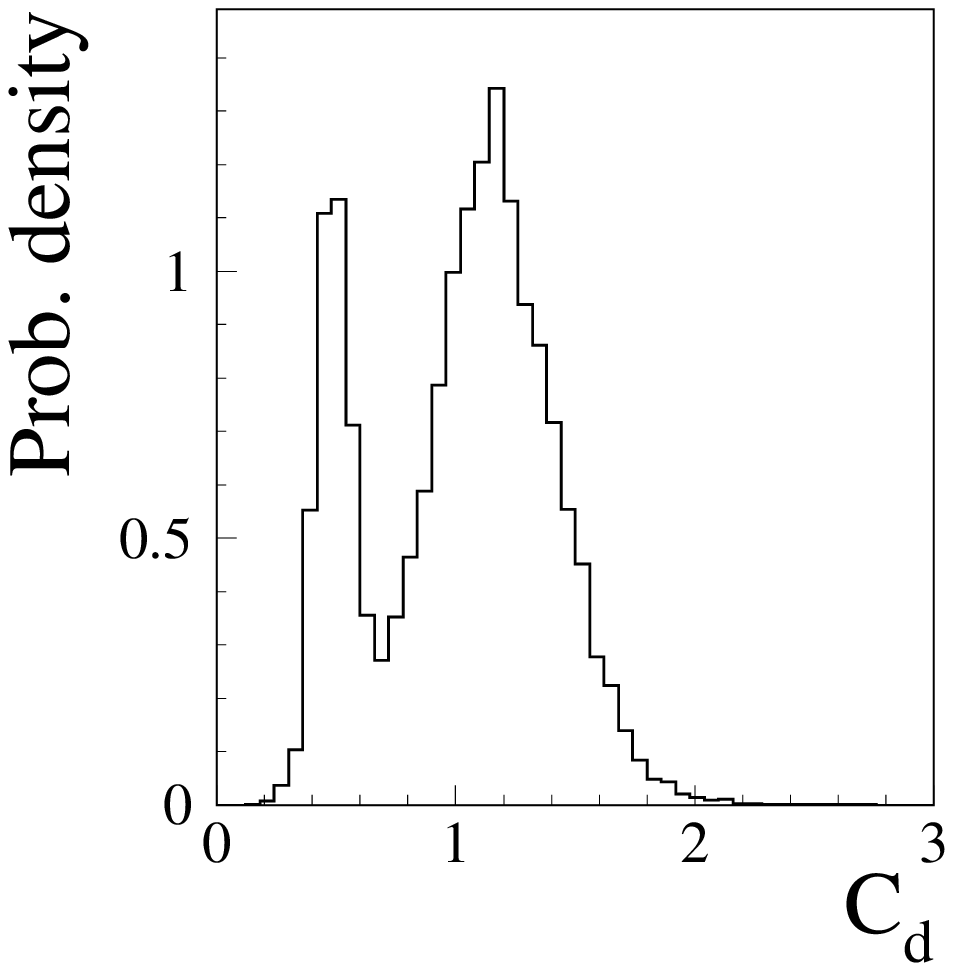}
\hss
\includegraphics[width=0.4\hsize]{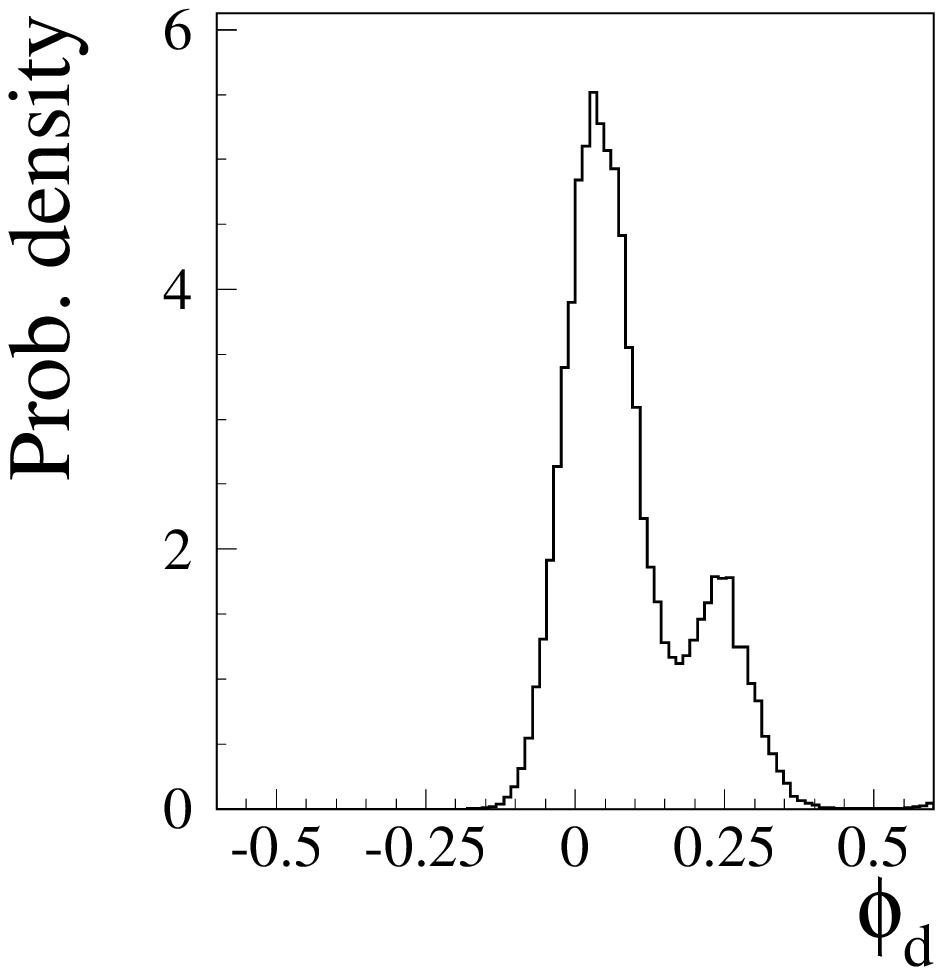}
\hss}
\hbox to\hsize{\hss
\includegraphics[width=0.7\hsize]{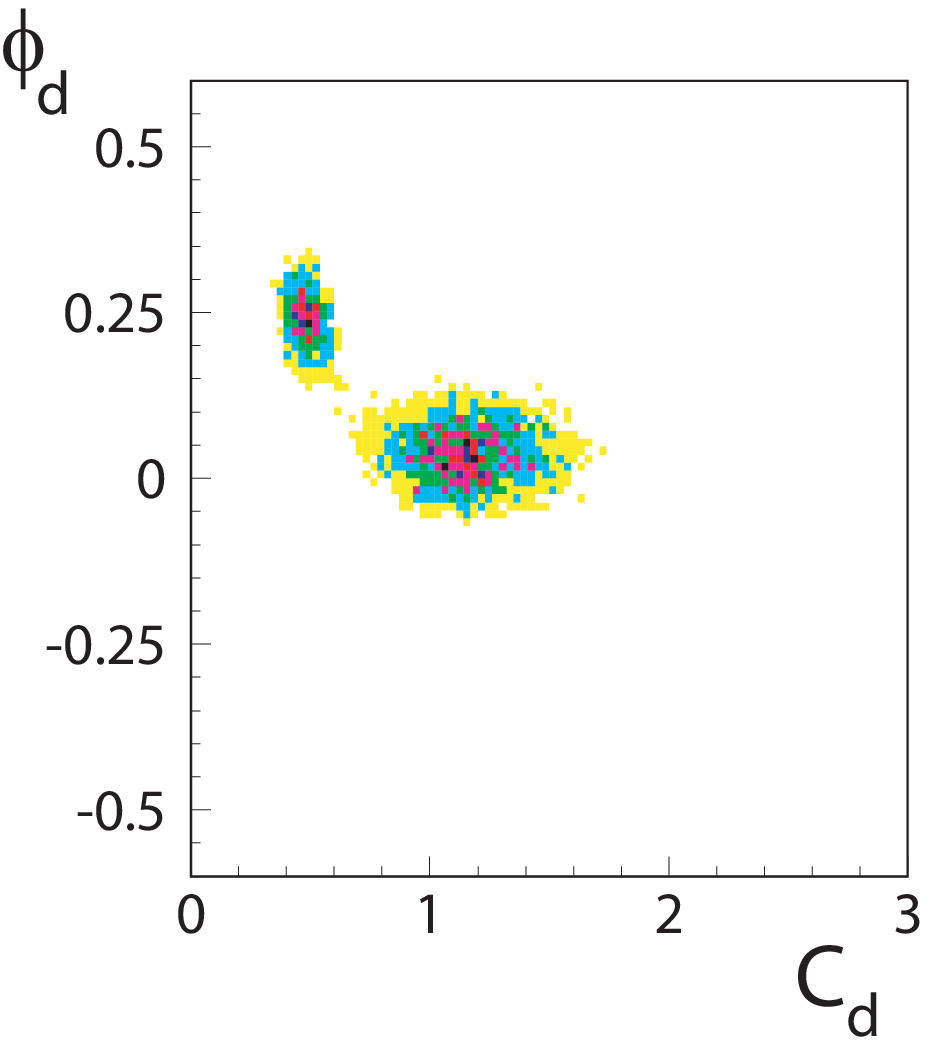}
\hss}
\caption[]{\it{The $C_d$ and $\phi_d$ distributions obtained by leaving these 
parameters free in the fit and the SM parameterization for $\left| V_{ub}\right|
/\left | V_{cb} \right|$ and $\epsilonk$ and $\Delta m_s$.}}
\label{fig:cd}
\end{figure}

Finally we explore the possibility of NP contributions in $B^0_d-\bar{B}^0_d$ 
mixing. Figure \ref{fig:cd} shows the distributions of $C_{d}$ and $\phi_d$.%
\footnote{The angle $\phi_d$ is determined from the measurement of $A_{CP}
(J/\psi K_s)$ up to a discrete ambiguity, $\phi_d+\beta \to \pi-\phi_d-\beta$.
In Figure \ref{fig:cd} only one of the two determinations (the one containing 
$\phi_d=0$) is shown for clarity.}

\begin{figure}[thb!]
\hbox to\hsize{\hss
\includegraphics[width=\hsize]{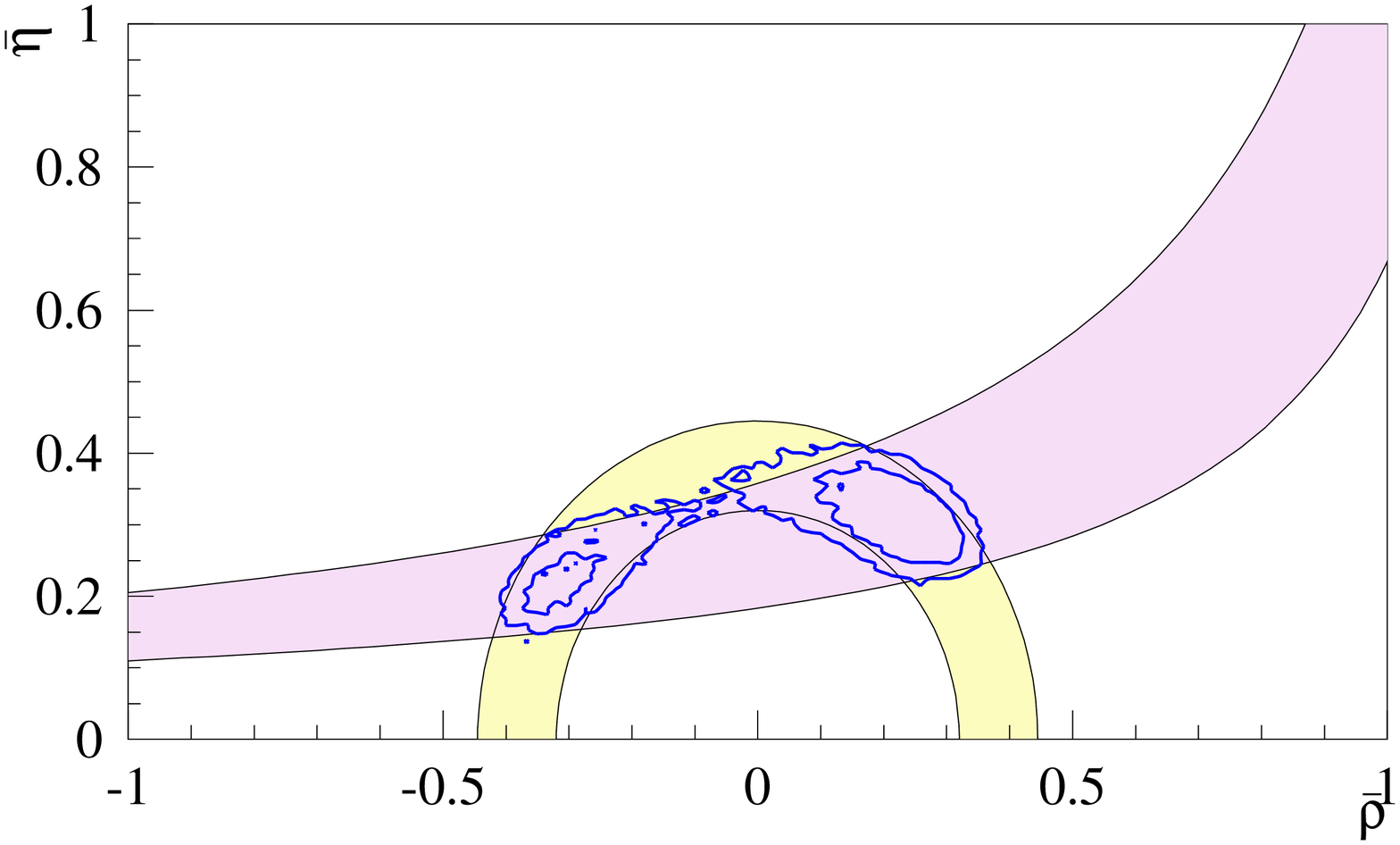}
\hss}
\caption[]{\it{The allowed regions for $\rhobar$ and $\etabar$ (contours at 
68\% and 95\%) as selected the measurements of $\left|V_{ub}\right|/\left|
V_{cb}\right|$ and $\epsilonk$. The bands at 95\% C.L. from the two constraints
are also shown.}}
\label{fig:cd2}
\end{figure}

\begin{figure}[thb!]
\hbox to\hsize{\hss
\includegraphics[width=0.4\hsize]{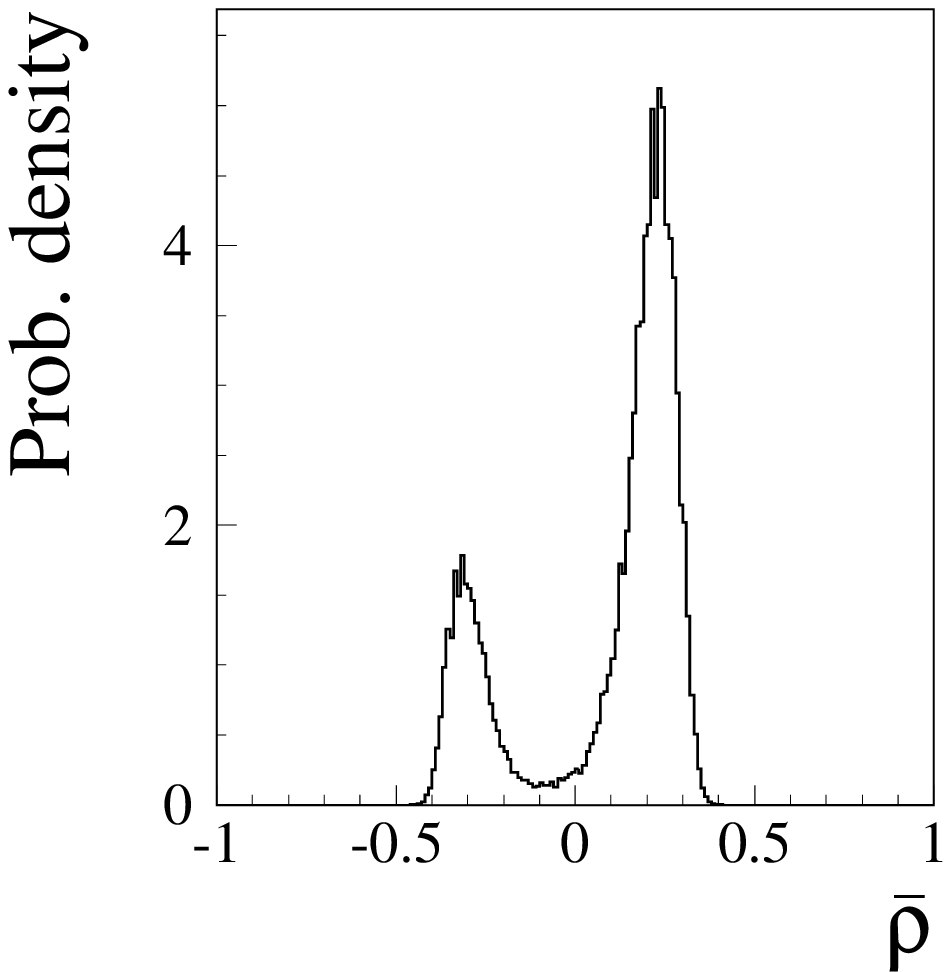}
\hss
\includegraphics[width=0.4\hsize]{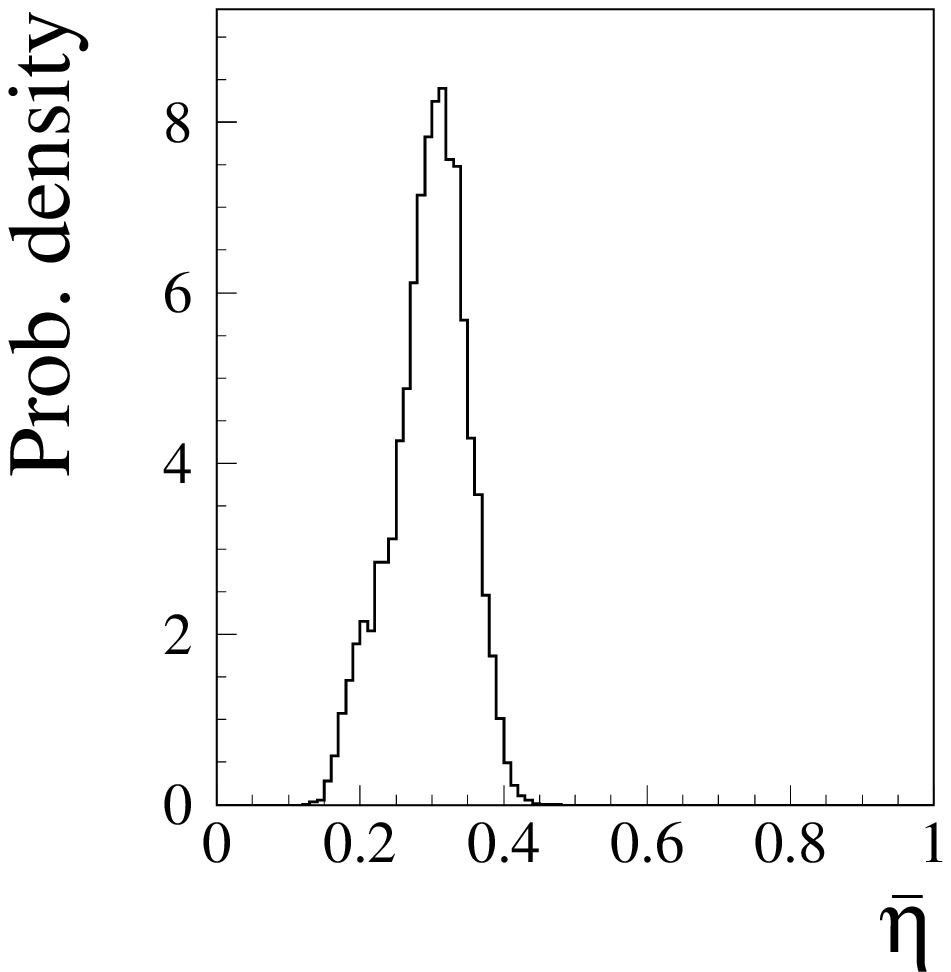}
\hss}
\hbox to\hsize{\hss
\includegraphics[width=0.4\hsize]{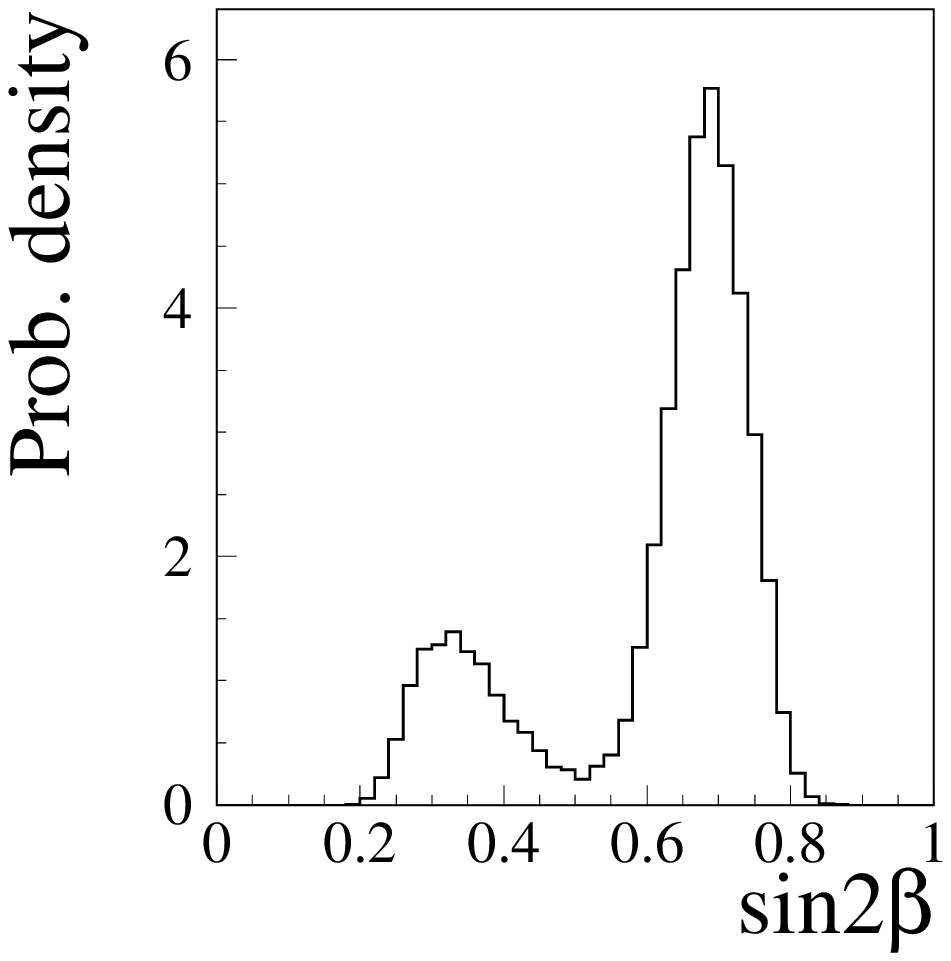}
\hss
\includegraphics[width=0.4\hsize]{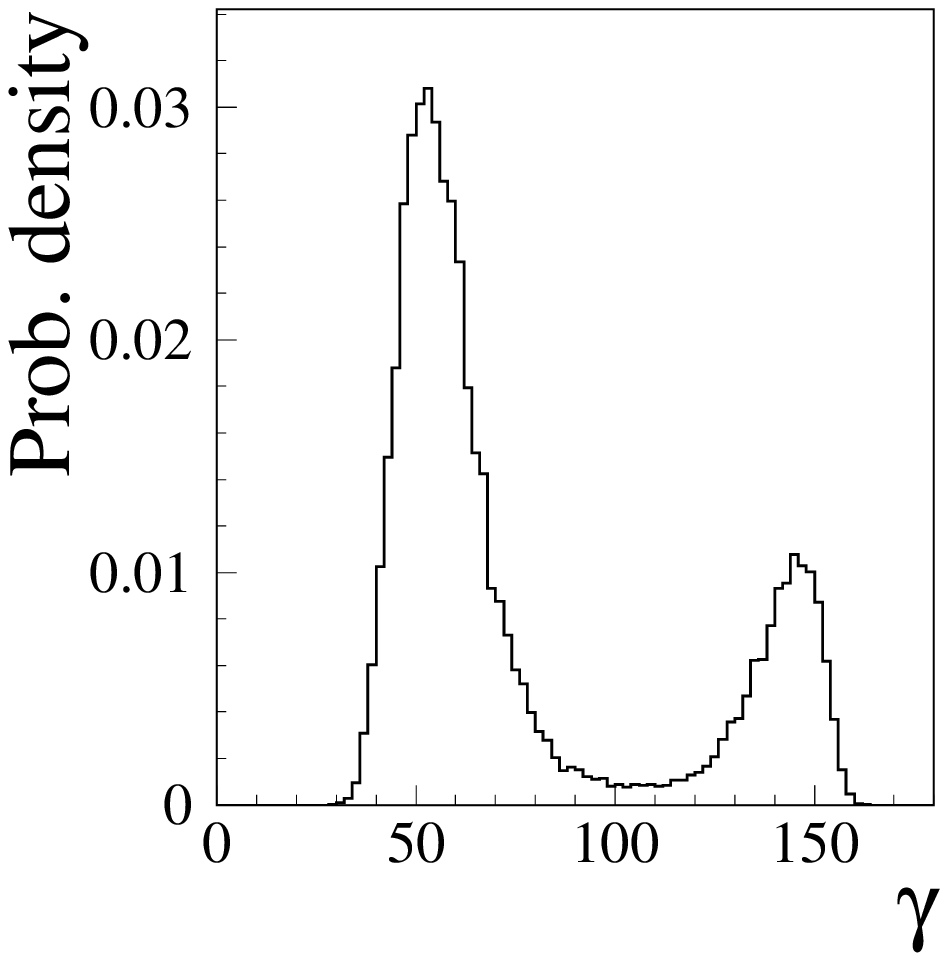}
\hss}
\caption{\it {The p.d.f. for $\rhobar$, $\etabar$, $\sin(2\beta)$ and $\gamma$ 
obtained when the NP parameters $C_d$ and $\phi_d$ are left free in the fit and 
the SM parameterization for $\left | V_{ub} \right |/\left | V_{cb} \right |$ 
and $\epsilonk$ are used.}}
\label{fig:1d_angle_cd}
\end{figure}

We see that two solutions are possible in this case. The first one peaks around 
the SM solution, $C_d\simeq 1$ and $\phi_d\simeq 0$. The second one represents 
instead the possibility of a really distinct NP contribution. The presence of 
two solutions corresponds to the fact that, as shown in Figure \ref{fig:cd2}, 
the constraint from $\epsilonk$ intercepts the circle defined by $\left| V_{ub} 
\right|/\left| V_{cb} \right|$ in two regions. 
The region on the quadrant with positive value of $\bar{\rho}$ corresponds to 
the SM solution. The presence of these two solutions is also visible in several 
distributions of the UT parameters shown in Figure \ref{fig:1d_angle_cd}.
As can be seen from these plots, in order to discriminate between the two
solutions, an independent determination of either $\rhobar$, as obtained for
instance from the study of the ratio $\Gamma(B\to K^*\gamma)/\Gamma(B\to \rho
\gamma)$ of radiative $B$ decays, or $\gamma$ would be necessary. The NP solution
has been also recently discussed in the literature~\cite{Isidori:2003ij}.

We want to conclude this discussion with a comment concerning the ``naturalness"
of this ``NP solution". First, we observe that NP, though possible, is certainly
not required in order to explain the results of the UT analysis. In addition,
suppose that the NP solution in $B^0_d-\bar{B}^0_d$ mixing is indeed the correct
one. Then it is somewhat surprising to find that in the SM, i.e. in the ``wrong"
theory of $B^0_d-\bar{B}^0_d$ mixing, the predicted values of $\dmd$ and
sin2$\beta$ actually select in the ($\rhobar$, $\etabar$) plane just one of the 
two regions which are also allowed by the other constraints, namely $\left| 
V_{ub} \right|/\left| V_{cb} \right|$ and $\epsilonk$. In this sense, the
consistency of this NP solution would be accidental.

\end{document}